\documentclass[12pt,preprint]{aastex}  \newcommand{\figfour}{0.7}

\newcommand{\nhh}{$N_{\mathrm{H\;\scriptscriptstyle{I}}}$\relax}

\begin{document}
 
\title{Origins of the \onequarter\,keV Soft X-Ray Background}

\author{Eric C. Bellm\altaffilmark{1} and John E. Vaillancourt\altaffilmark{2}}
\altaffiltext{1}{current address: Harvard-Smithsonian Center for Astrophysics, 60 Garden St., Cambridge, MA 02138; bellm@fas.harvard.edu}
\altaffiltext{2}{current address: University of Chicago, Enrico Fermi Institute, 5640 S. Ellis Ave., Chicago, IL 60637; johnv@oddjob.uchicago.edu}
\affil{Physics Department, University of Wisconsin, 1150 University Ave., Madison, WI 53706}

\begin{abstract}
Snowden and coworkers have presented a model for the \onequarter\,keV
soft X-ray diffuse background in which the observed flux is dominated
by a $\sim 10^6$\,K thermal plasma located in a 100--300 pc diameter
bubble surrounding the Sun, but has significant contributions from a
very patchy Galactic halo.  Halo emission provides about 11\% of the
total observed flux and is responsible for half of the \ion{H}{1}
anticorrelation.  The remainder of the anticorrelation is presumably
produced by displacement of disk \ion{H}{1} by the varying extent of
the local hot bubble (LHB\@).  The \emph{ROSAT} R1 and R2 bands used
for this work had the unique spatial resolution and statistical
precision required for separating the halo and local components, but
provide little spectral information.  Some consistency checks had been
made with older observations at lower X-ray energies, but we have made
a careful investigation of the extent to which the model is supported
by existing sounding rocket data in the Be (73--111 eV) and B bands
(115--188 eV) where the sensitivities to the model are qualitatively
different from the \emph{ROSAT} bands.  We conclude that the
two-component model is well supported by the low-energy data.  We find
that these combined observations of the local component may be consistent with
single-temperature thermal emission models in collisional ionization
equilibrium if depleted abundances are assumed. However, different
model implementations give significantly different results, offering
little support for the conclusion that the astrophysical situation is
so simple.  
\end{abstract}


\section{Introduction} 

Early observations of the \onequarter\,keV diffuse X-ray background
showed a strong anticorrelation with Galactic \ion{H}{1} column
density that suggested an extragalactic source absorbed by foreground
interstellar gas. However, it gradually became apparent that a) there
is a large flux in the Galactic plane, even though unit optical depth
at \onequarter\,keV is only $1 \times 10^{20}$ H atoms cm$^{-2}$, b)
the apparent absorption cross sections observed at intermediate and
high Galactic latitudes are much smaller than the calculated
photoelectric values, and c) the still softer B and Be bands show, on
the average, almost identical variations with $N_\mathrm{H}$ despite
absorption cross sections that should be 2--10 times larger (Table
\ref{tbl-tau}). These three observations were readily explained by a
model in which the \onequarter\,keV flux originates in an irregular
local hot bubble (LHB) of 100--300 pc extent surrounding the Sun
\citep{sanders77,xrb_review}.  In this model, the X-ray/\ion{H}{1}
anticorrelation is produced by the displacement of neutral material by
a $10^6$\,K plasma.  \citet{snowden90} showed that \ion{H}{1} column
densities were consistent with the excavation of a standard
plane-parallel disk by a uniform X-ray emitting bubble of extent
varying as required by the observed X-ray brightness.  With the
limited statistics available, the small variations in the X-ray band
ratios appeared random and uncorrelated with \nhh, as would be
expected for this model with all emission in the foreground.

\begin{figure}
\plotone{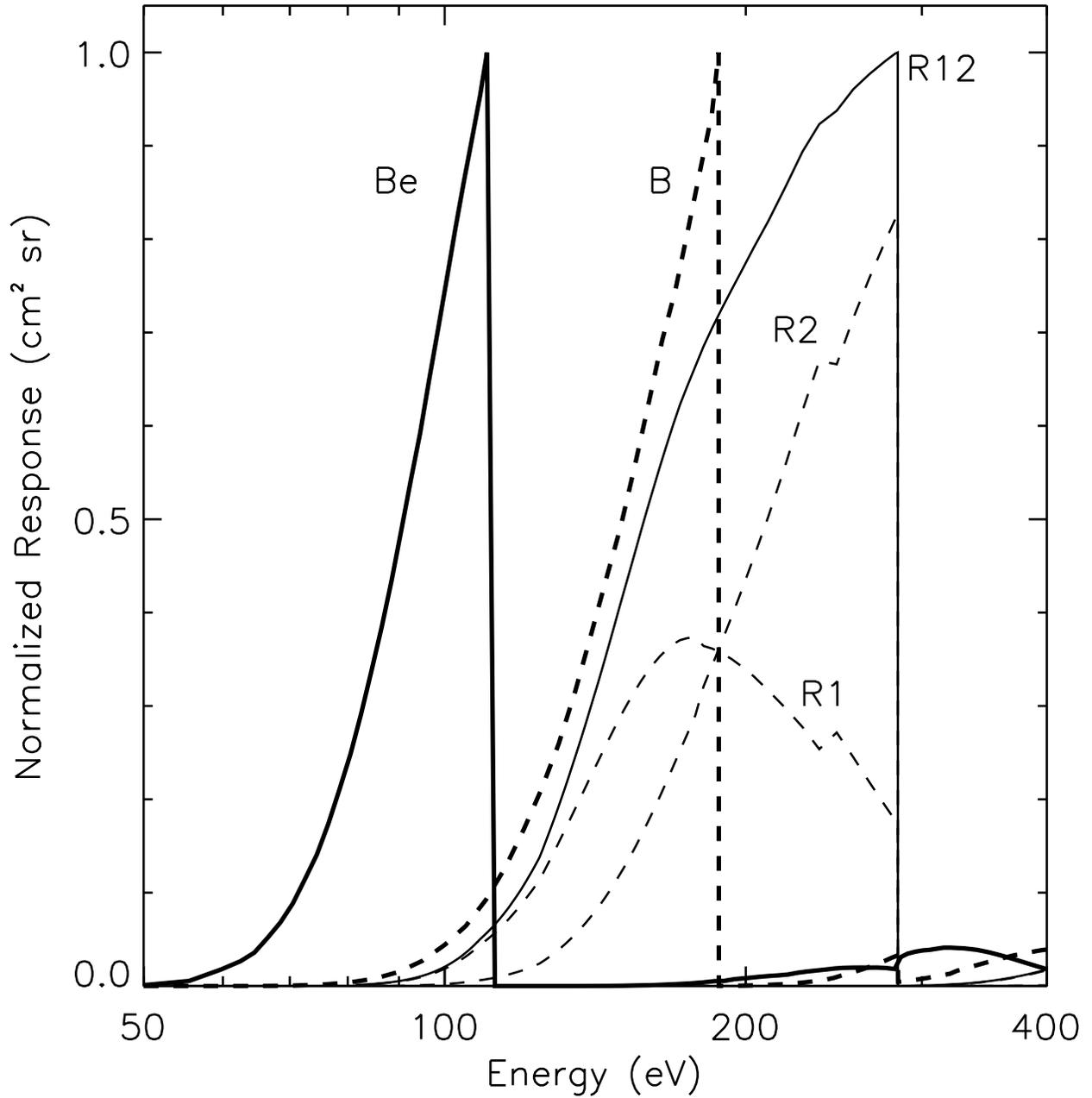}
\caption{Normalized response functions for the pulse-height bands
considered in this work.  The B and Be bands are from the sounding
rocket investigations of \citet{mccammon83}, \citet{bloch86},
\citet{edwards}, and \citet{juda91}. The R1, R2, and R12 ($\equiv$ R1
+ R2) bands are from the \emph{ROSAT} satellite
\citep{snowden94,snowden97}; these three bands are normalized to the
peak of the R12 band.
\label{fig-response}}
\end{figure}

Some of the first diffuse background observations utilizing the improved angular resolution 
of \emph{ROSAT}, however, revealed shadowing of half the observed
\onequarter\,keV X-ray flux toward molecular clouds known to lie
several hundred parsecs above the Galactic plane, showing that a
significant part of the \onequarter\,keV emission is from above the
Galactic disk (e.g. \citealt{burrows91};
\citealt{snowden91,ursamajor}). To account for this,
\citeauthor{snowden98}\ (\citeyear{snowden98}, hereafter S98) proposed
a multi-component model that includes both a LHB and patchy halo
emission absorbed by the intervening disk gas (see also
\citealt{kuntzsnowden}). They have estimated the halo contribution by
fitting spatial variations in the R1 and R2 bands of the \emph{ROSAT}
all-sky survey to variations in the \ion{H}{1} column density on
angular scales too small for displacement to have an appreciable
effect. The remainder is assumed to come from the LHB as before with
no absorption. We refer to the latter part as the local R12 rate (R12
band = R1 + R2; see Fig.\ \ref{fig-response}).  A fixed contribution
from extragalactic point sources was also included, described by a
power-law in energy.

S98 capitalized on the improved angular resolution ($\sim$$12\arcmin$) and very good
statistical precision of the \emph{ROSAT} all sky survey to isolate
the effects of absorption and displacement and allow division of the
observed flux into two components.  However, the R1 and R2 bands have
effective energies that differ by only a small fraction of their
widths and provide limited spectral information
(Fig.\ \ref{fig-response}; Table \ref{tbl-tau}).  A $10^6$\,K plasma
radiates strongly at lower energies as well, where the interstellar
absorption cross sections are much larger.  As yet there have been no
useful satellite surveys at lower energies and we must therefore rely
on existing data from older sounding rocket surveys, despite their
much poorer angular resolution and statistical precision.  A series of
ten rocket flights from 1972--1982 completed an all-sky survey in the
B band (115--188 eV), as well as a C band equivalent to the
\emph{ROSAT} R12 band \citep{mccammon83}. 
Additional sounding rocket flights between 1984--88 made a total of 25
pointed observations at soft X-ray energies in the Be band
\citep[77--111 eV;][]{bloch86,edwards,juda91}. With their limited
statistical precision, both sets of observations were consistent with
the total flux in these bands originating in the LHB\@.  However, the
division of R12 intensity into local and halo components makes very
specific predictions for the behavior of the softer bands.  S98
checked for consistency with the Be band observations, but in this
paper we investigate whether a more complete analysis with all of the
available low energy data adds support for their model.

In \S\S\ \ref{sec-local} and \ref{sec-distant} we compare the behavior
in the B and Be bands to what would be expected from the specific
halo/local division found by the analysis of S98. In \S
\ref{sec-spectra} we investigate whether all the observed band ratios
(B/R12, Be/R12, and R1/R2) can be fit by a single-temperature emission
model.  We find that with current collisional ionization equilibrium
models, depleted abundances are necessary to achieve even rough
consistency with the observed band ratios. We also show that the
fitted model parameters depend strongly on the particular model.

\section{Isolating Local Emission} \label{sec-local} 

At Be band energies, almost complete extinction is produced by even
the smallest observed column densities of neutral hydrogen
($\sim$$10^{19}$ cm$^{-2}$ for unit optical depth; Table
\ref{tbl-tau}). Since all Be band measurements were made towards
regions with $N_\mathrm{H}=1$--$6\times 10^{20}$\,cm$^{-2}$
\citep*{sfd98} virtually all Be band counts must originate from the
LHB\@.  The smaller cross section for R12 photons (Table
\ref{tbl-tau}) means that a larger fraction of the counts in that band
may be non-local. \citet{juda91} found only negligible column
densities ($\sim10^{18}$ cm$^{-2}$) of $N_\mathrm{H}$ intermixed in the
LHB in their best fit to the Be band data.

Accordingly, the ratio of the Be band counts to R12 counts produced
locally should be constant for uniform conditions in the LHB\@.  As
noted by S98, their modeled local R12 rates display this expected
correspondence with the Be band observations, but similar conclusions
were reached using the total \onequarter\,keV flux during early
analysis of the Be band data, which had supported the LHB-only model
\citep{bloch86}.  To make a direct comparison, we obtained total
\citep{snowden97} and local (S98) R12 band rates in the directions of
the 25 Be band measurements by averaging over the $15\arcdeg$ FWHM
fields of the rocket observations.  We have excluded directions which
overlap atypical features in the X-ray maps such as the Eridanus
enhancement, the Mono-Gem Ring, the Cygnus superbubble, the Vela SNR,
and the North Polar Spur (see \citealt{rosatmaps1}).  We have also
excluded a region of enhanced $N_\mathrm{H}$ where there exists
evidence for Be band absorption \citep[$l=132\fdg6$,
  $b=-76\fdg8$;][]{edwards, burrows84}.

\begin{figure*}
\plotone{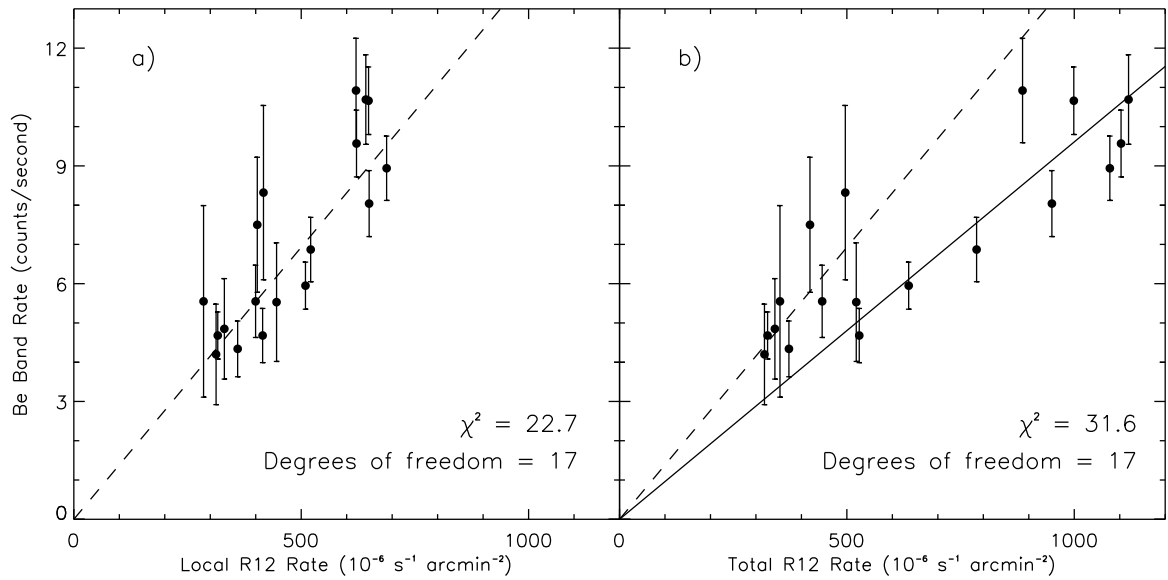}
\caption{ a) Wisconsin Be band counts vs.\ local R12 rate predicted by
  the model of \citet{snowden98}.  The best fit slope of 0.0139 is
  shown as a dashed line.  b) Wisconsin Be band counts vs.\ total R12
  rate.  The best fit slope of 0.0096 is shown as a solid line.  The
  dotted line is the best fit slope from (a).  Regions of the sky with
  atypical X-ray flux features have been excluded (see text).
\label{fig-bevsr12} }
\end{figure*}

Comparing the plots in Figures \ref{fig-bevsr12}a (Be vs.\ local R12)
and \ref{fig-bevsr12}b (Be vs.\ total R12) we see that the Be rates
have a noticeably better linear correspondence with just the local
part of the R12 rates than they do with the total R12 rates.  That is,
a linear fit to the Be vs.\ R12 rates yields a smaller $\chi^2$ for
the local R12 rates than for the total R12 rates. The Be to total R12
ratios are consistently low in just those directions where there is a
large halo contribution in the S98 two-component model.

\section{Correlations with Column Density} \label{sec-distant}

\subsection{Energy-dependence of Absorption}

The two component model also makes definite predictions for changes in
the spectral distribution of the distant component with \nhh\ 
due to energy-dependent absorption.  We estimated observed intensity
of the halo emission in the R1, R2, R12, and B bands by subtracting
the local rates in each band from the total observed rate. The local
R1 and R2 rates were taken from the S98 fits.  For the B band, the
local R12 map was first scaled by the observed low-latitude B/R12
ratio ($\sim 0.11$; with B band units of counts s$^{-1}$ and R12 band
units of $10^{-6}$ counts s$^{-1}$ arcmin$^{-2}$) to obtain an
estimate of the local B band rate. This local rate was subtracted from
the observed total B band map of the Wisconsin sky survey
\citep{mccammon83} to determine observed B band counts ostensibly from
the halo.  For this study we have binned the all-sky data into
$15\arcdeg \times 15\arcdeg$ sections and limited the observations to
those at high Galactic latitudes ($|b| > 45\arcdeg$), as the lower
latitude points are dominated by LHB emission. We have also excluded
the atypical regions discussed in \S\ref{sec-local} and regions where
the B band was contaminated by electrons or X-rays produced in the
solar wind or terrestrial atmosphere \citep{mccammon83}.

For S98's assumed uniform halo emission spectrum, the R1/R2 and B/R12
band ratios for the observed halo component should vary in a
predictable way with absorbing column density. The hydrogen column
density in each direction was determined by scaling the $100\,\micron$
fluxes of \citet{sfd98} by a factor of $1.475 \times
10^{20}$\,cm$^{-2}$\,MJy$^{-1}$\,sr (determined by
\citealt{kuntzsnowden}) and averaging over the sky bin.

\begin{figure}
\epsscale{\figfour}
\plotone{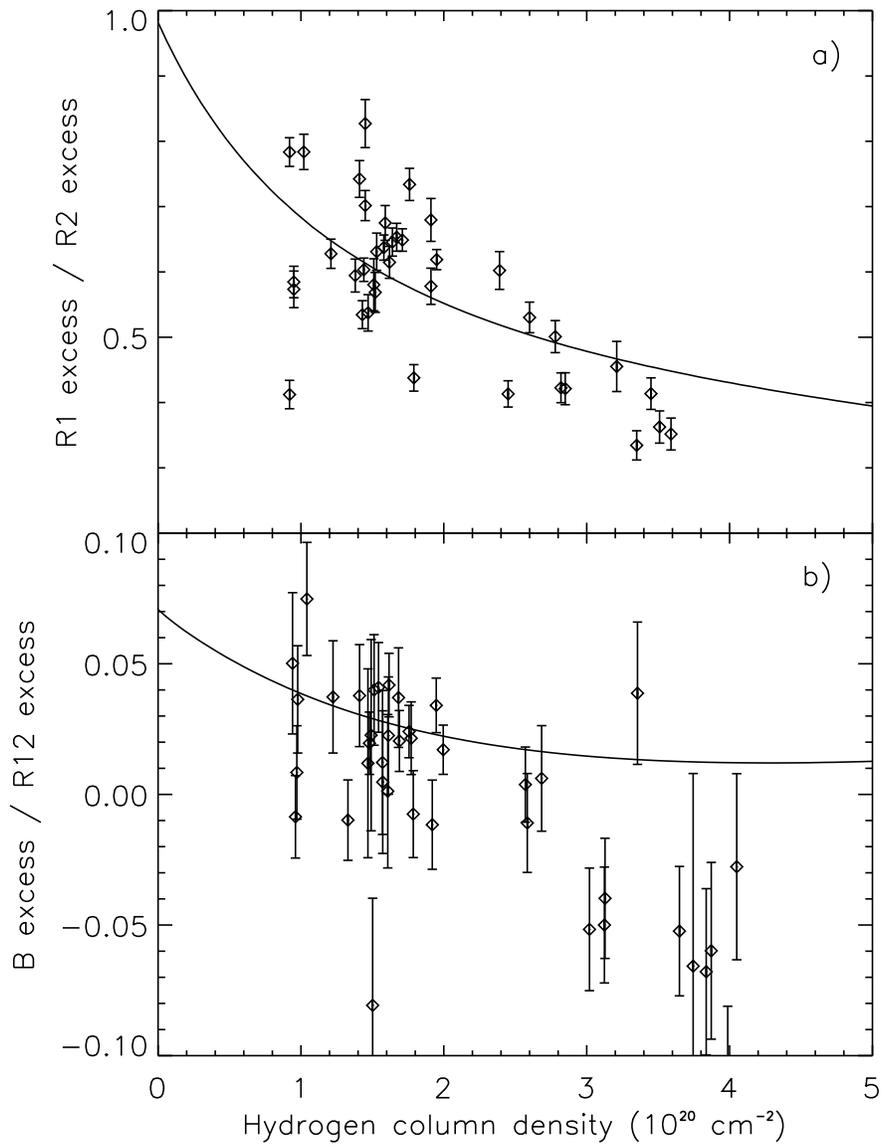}
\caption{Band ratios of absorbed halo emission inferred from the
  \citet{snowden98} model vs.\ neutral hydrogen column density.
Data at low Galactic latitudes ($|b| < 45\arcdeg$) have been excluded.
The ratio predicted by the \citeauthor{snowden98}\ model is shown by
the solid curves.
Absorbed count rates were obtained by subtracting the modeled local
count rates from the total observed rates for $15\arcdeg\times
15\arcdeg$ bins on the sky.  a) Ratio of the \emph{ROSAT} R1 to R2
band.  b) Ratio of B band to R12 band.
The error bars in (b) take into account only the uncertainties of the
B band measurements; the R12 uncertainties are small compared to those
in the B band.  The barely perceptible rise in the model curve at high
$N_\mathrm{H}$ in (b) is due to a ``leak'' in the B band response at
$\sim 400$\,eV which is larger than that in the R12 band (see
Fig.\ \ref{fig-response}).
\label{fig-excess_ab}}
\end{figure}

Figure \ref{fig-excess_ab} shows the band ratios for these fields
versus \ion{H}{1} column density. We have also plotted the ratios
expected from \citet{raymondsmith}'s plasma emission model at a plasma
temperature of $\log T = 6.02$ (S98's modeled halo temperature). (This
model was implemented in
XSPEC\footnote{\url{http://heasarc.gsfc.nasa.gov/docs/xanadu/xspec/}}
v11.2 [e.g.\ \citealt{xspec1}]; see \S\ref{sec-spectra}).  These plots
are very ``noisy'' because subtraction of the large local component
leaves the residual dominated by statistical errors and by small
deviations from the uniform assumptions of the model.  Nonetheless,
the ratios do display the expected trends with column density and are
not consistent with the LHB-only model, where the thickness of the
disk \ion{H}{1} should have no effect on the spectrum.

As a corollary to this, the unabsorbed intensity of halo emission
should be independent of foreground column density.  This is a rather
strong test, since the halo intensities are computed by dividing the
observed halo component by the calculated gas transmission, a large
correction that is itself directly correlated with column density.
\citet{kuntzsnowden} performed this test and their Figure 4 appears
remarkably uncorrelated.

\subsection{Relative Local/Halo Flux Contributions}
Due to the large interstellar absorption, the halo component
contributes only about 11\% to the average observed intensity in the
R12 band.  However, in a few directions where the column density is
low and the halo happens to be bright, it dominates the total
emission, and it must contribute strongly to the observed
anticorrelation with \ion{H}{1}\@.  In Figure \ref{fig-r12compare} we
plot the rates in the \emph{ROSAT} R12 band vs.\ $N_\mathrm{H}$ (again
the data have been binned in $15\arcdeg \times 15\arcdeg$ sections on
the sky).  The data points labeled ``local'' are the LHB emission as
modeled by S98, while the ``observed halo'' rates are the actual total
rates minus the local rates.  Both are strongly anticorrelated with
\ion{H}{1} column density.  For the halo component, this should be due
to absorption, and for the LHB component is presumably due to
displacement.  The two effects can be seen to contribute about equally
to the observed anticorrelation.

\subsection{Solar Wind Charge-Exchange}
\begin{figure}
\epsscale{1.0}
\plotone{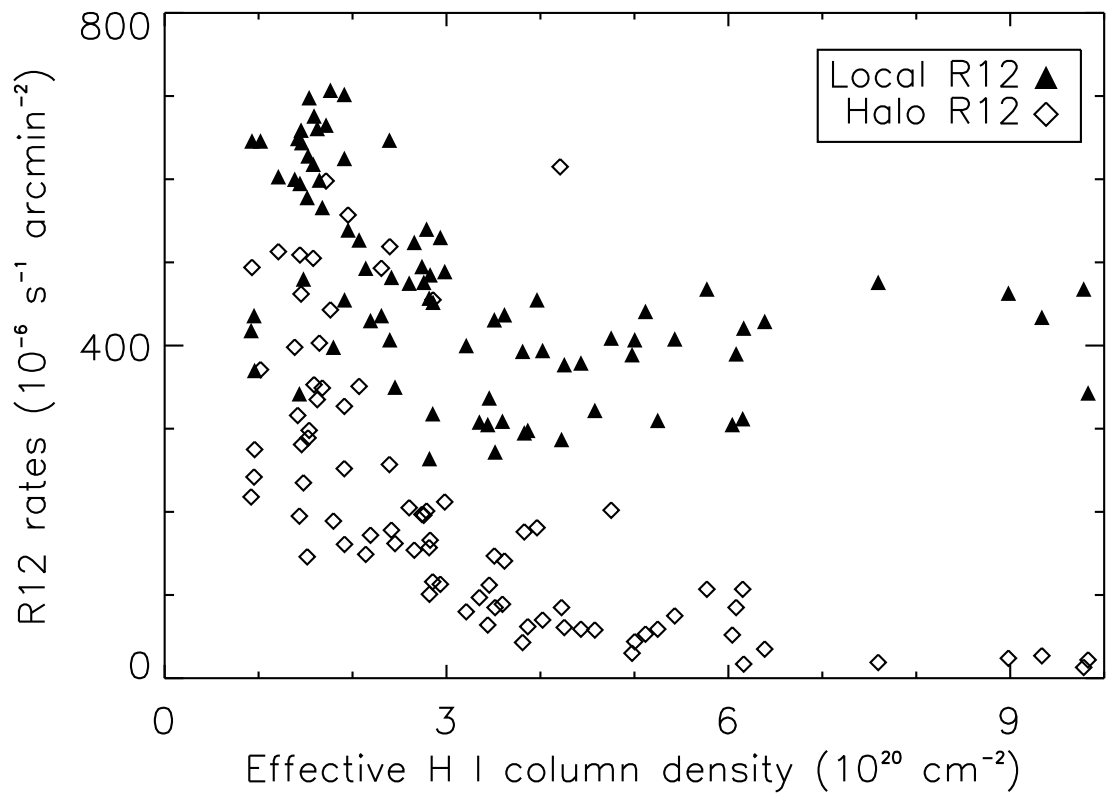}
\caption{Local Hot Bubble \citep{snowden98} and observed halo
  ($\equiv \mathrm{total} - \mathrm{LHB}$) rates in the
  \emph{ROSAT} R12 band vs.\ column density.  Both components
  contribute about equally to the observed X-ray - \nhh\
  anticorrelation.  This also implies that at least some of the LHB
  emission is not local to the solar system.  Error bars are
  approximately the same size as the data points.  All Galactic
  latitudes are included in this plot but we have excluded the same
  contaminated and atypical regions of the sky as in Figures
  \ref{fig-bevsr12} and \ref{fig-excess_ab}.
\label{fig-r12compare}}
\end{figure}

The anticorrelation of local R12 with $N_\mathrm{H}$ also provides a
constraint on the portion of the local flux which may be attributed to
solar wind charge-exchange emission (CXE\@).  Since the CXE should not
be correlated with the Galactic disk \ion{H}{1} column density, the
\nhh-dependent portion in Figure \ref{fig-r12compare} implies a
non-zero contribution to this flux from a LHB\@.  The magnitude of the
\nhh-independent portion is consistent with the results of
\citet{lallement04a} who puts an upper limit of $\sim 300\times
10^{-6}$\,counts s$^{-1}$ arcmin$^{-2}$ on CXE\@.  
Since the all-sky average of the S98 local component is 
$455\times 10^{-6}$ counts s$^{-1}$ arcmin$^{-2}$ (Figure \ref{fig-r12compare}),
at least 34\% of it must originate in an extensive LHB.



\section{Constraints on Spectral Models} \label{sec-spectra}

In a more sophisticated analysis of the S98 model,
\citeauthor{kuntzsnowden} (2000) derived an emission temperature of
$\sim 10^{6.1\pm 0.1}$\,K for the LHB\@.  The halo component was
consistent with the same temperature but had larger variations.  For
the halo emission, the small contribution to the observed brightness
in the B and Be bands means that there is little information to be
added by using these older data.  However, in the LHB the low energy
data add a considerable amount of spectral information.  Fits to each
of these bands to thermal emission models have been done before
\citep[e.g.][]{dxs}, but it is
worthwhile to revisit the issue utilizing newer spectral models.

\begin{figure}
\epsscale{\figfour}
\plotone{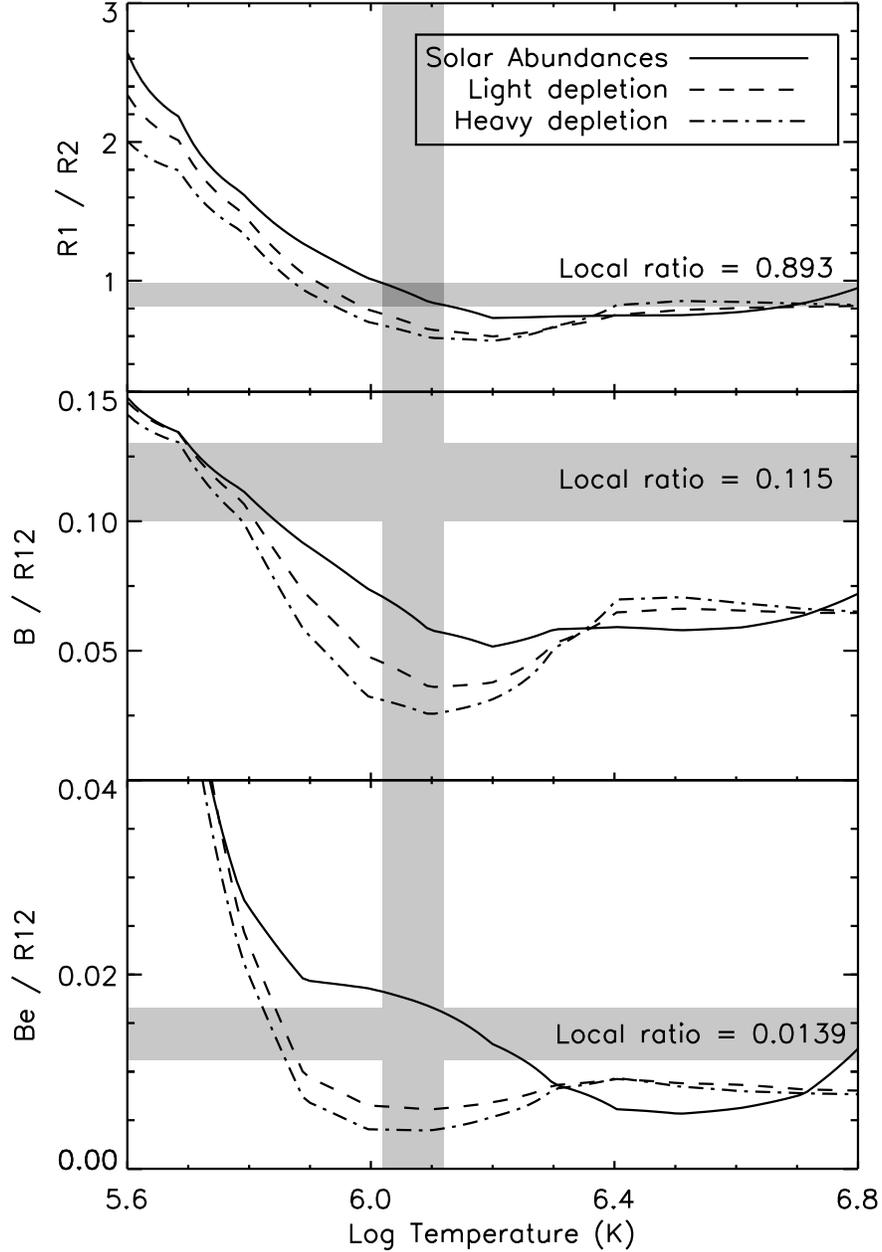}
\caption{Band ratios as a function of plasma temperature. We have used
  the collisional ionization equilibrium model of \citet{raymondsmith}
  (implemented in XSPEC v11.2) to determine the ratio of X-ray
  emission in the bands in Figure \ref{fig-response}.  We compare
  three sets of elemental abundances tabulated by
  \citet{savagesembach}: solar abundances, and those observed towards
  warm (light depletion) and cool (heavy depletion) clouds.  The
  ratios observed at low Galactic latitudes are shown as horizontal
  gray bands; the widths represent only estimates of the measured
  ratio's distribution but are not formal errors. The local R1/R2
  ratio is from S98, the B/R12 and Be/R12 (see
  Fig.\ \ref{fig-bevsr12}) ratios are from this work.
The vertical gray band centered at $\log T = 6.07$ indicates the best
fit temperature to the R1/R2 ratio obtained by S98 using the
 \citeauthor{raymondsmith} model with solar abundances.  
\label{fig-bandvst}}
\end{figure}

At low galactic latitudes, large absorbing columns ensure that all
counts observed at \onequarter\,keV and below are generated in local
source material. From low latitude observations we estimate B/R12 =
0.115.  From the fit of Be band counts to S98's local component, we
find Be/R12 = 0.0139 (see Figure \ref{fig-bevsr12}a).  (This Be/R12
ratio is consistent with the value obtained by calculating the ratio
of Be to total observed R12 for the limited number of available low
latitude fields.)  Figure \ref{fig-bandvst} compares these ratios with
the single-temperature collisional ionization equilibrium (CIE) model
of \citeauthor{raymondsmith}, implemented in Xspec v11.2.
(This version is consistent with that used by S98.)  Assuming solar
abundances, this model gives quite different temperatures for the
three band ratios.  We find $\log T = 5.8$ for B/R12 and 6.2 for
Be/R12, whereas S98 found $\log T = 6.1$ from the R1/R2 ratio (see
Fig.\ \ref{fig-bandvst}; Table \ref{tbl-spectra}).

Figure \ref{fig-bandvst} also shows the predicted band ratios for
models in which the heavy elements have been depleted with respect to
their solar abundances, again using the Raymond \& Smith CIE
model. The depleted models use abundances observed in a warm cloud
(light depletion) and a cold cloud (heavy depletion) toward $\zeta$
Ophiuchi \citep{savagesembach}.  None of these models yields entirely
consistent temperatures for all three band ratios (see Table
\ref{tbl-spectra}), but the depleted abundances fare much better than
solar, with the heavy depletion model giving the smallest scatter and
a temperature near $\log T = 5.84$.  One might think that all dust
grains should be evaporated in a $10^6$\,K plasma, and that it is
therefore reasonable to consider only solar abundances.  However,
calculations of grain destruction in shocks and by hot gas sputtering
show that over the broad range of feasible histories for the LHB, the
surviving fraction of silicate dust mass ranges from $>90\%$ for solid
grains after $10^6$ years, down to 30\% for fluffy grains after $10^7$
years (\citealt{smith96} and references therein).  Initial depletions
onto the silicate grains are well over 95\% for Si and Mg, which are
the important emitters in the R12 and B bands, and about 99\% for Fe,
which dominates the Be band, so final depletions are effectively just
the surviving fraction of dust mass.  An additional complication of
dust is that while sputtering is taking place, cooling is enhanced and
the spectrum altered by lines of the under-ionized atoms recently
added to the gas phase.

Bragg crystal spectrometer measurements at low galactic latitudes with
considerably better energy resolution found that the LHB emission was
not acceptably fit by either single-temperature CIE or by any
reasonable non-CIE model, but that moderate to heavy depletions of Si
and Mg at least offered considerable improvement \citep{dxs}.
Observations with microcalorimeters at high latitudes produced
qualitatively similar results \citep{rocketpaper}.  A recent high
spectral resolution observation of the EUV diffuse background by the
CHIPS satellite has placed upper limits on the Fe lines near 72\,eV
that would require Fe depletion of at least $10\times$ to be
consistent with a single hot plasma as the source of the soft X-ray
flux in the R12 and B bands \citep*{hurwitz04}.  Their best fit
temperature is $\log T = 5.9$, in reasonable agreement with our
depleted abundance fits to the soft X-ray bands.
However, single-temperature CIE emission is an idealization of any
possible astrophysics and there are other likely contributors to the
observed X-rays, such as charge-exchange of solar wind ions in
interplanetary space \citep{lallement04a}, so the actual
situation is probably not this simple.  While all of these data are
consistent with an origin in a single hot plasma, they by no means
rule out a substantial contributation from interplanetary
charge-exchange.
 
\begin{figure}
\epsscale{\figfour}
\plotone{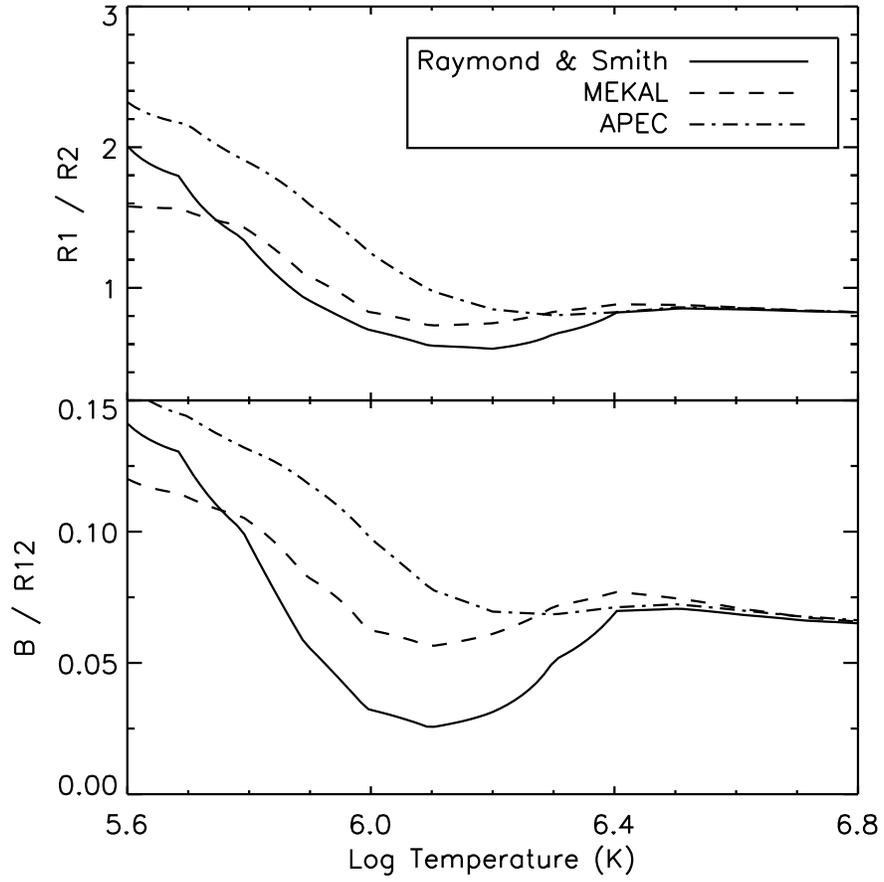}
\caption{Band ratios as a function of temperature for different models
  of equilibrium plasmas, all using heavily depleted elemental
  abundances (as in Fig.\ \ref{fig-bandvst}).
\label{fig-compare_models}}
\end{figure}

In Figure \ref{fig-compare_models} we compare band ratios calculated
with \citeauthor{raymondsmith} and two other widely used equilibrium
emission models: MEKAL \citep*{mekal1985,mekal1986} and
APEC\footnote{\url{http://cxc.harvard.edu/atomdb/sources\_apec.html}}
\citep{smith01}, both also as implemented in XSpec v11.2.  All used
the cold cloud (heavily depleted) abundances.  It can be seen that
temperatures derived for a given band ratio have a very significant
dependence on the model.

\section{Summary} 

Sounding rocket data in the low-energy B and Be bands add support for
the \citet{snowden98} model in which the source of the
\onequarter\,keV diffuse background is composed of both a local
unabsorbed component and a very patchy ``halo'' component absorbed by
neutral gas in the Galaxy.  Although the halo emission measure is
quite large in some directions, its contribution to the observed flux
is modest because of the heavy absorption in most directions.
Averaged over the sky, the halo component is responsible for only
10.7\% of the \emph{ROSAT} R12 band, 4.7\% of the B band, and 0.4\% of
the Be band. However, it supplies more than half the observed
intensity in selected directions and is responsible for about half the
anticorrelation with \nhh.  Displacement effects in the LHB presumably
account for the remainder.  The X-ray - \nhh\ anticorrelation in the
local R12 flux implies that at least 34\% of this flux must come from
hot gas in the LHB rather than charge-exchange reactions in the solar
wind.

We also find that \citeauthor{raymondsmith} collisional ionization
equilibrium models require depleted abundances to come close to
providing a single-temperature solution for the observed Be/R12 and
B/R12 band ratios in the local hot bubble.  A temperature of $\log T
\sim 5.84$ with heavy depletions is most consistent with both these
and S98's modeled R2/R1 ratio in the LHB, for which they had found
$\log T = 6.07$ using solar abundances.  However, other available CIE
emission models give quite different results, which does little to
lend confidence to what may be a fortuitous fit.  The very few
available observations of the $\sim\onequarter$\,keV diffuse
background with better spectral resolution also favor depleted
abundances, but suggest that the spectral picture is more complex than
single-temperature plasma emission.  More realistic astrophysical
situations and contributions from other sources such as solar wind
charge-exchange are likely to be parts of the picture.  Disentangling
and characterizing these will almost surely require measurements of
individual line intensities and physical analysis of their ratios
rather than modeling of broad band rates.

\acknowledgements We would like to thank Steve Snowden for providing
the maps of his model of the \onequarter\,keV flux and Kip Kuntz for
providing the \citeauthor{raymondsmith} spectrum used by S98. We would
also like to thank Dan McCammon and Wilt Sanders for useful
discussions, direction, and assistance with data analysis and
interpretation.  This work was supported by a NSF-REU site grant
(AST-0139563) to the University of Wisconsin-Madison and E. C. B., and
by NASA grant NAG5-5404.

\clearpage

\bibliography{mccammon,xray,dust}
\bibliographystyle{apj}

\clearpage

\begin{deluxetable}{lccc}
\tablecolumns{4}
\tablewidth{0pt}
\tablecaption{Band Cross Sections\label{tbl-tau}}
\tablehead{\colhead{Band} & \colhead{Energy\tablenotemark{a}} & \colhead{Cross Section\tablenotemark{b}} &
 \colhead{Reference}\\ \colhead{} & \colhead{(eV)} & \colhead{($10^{-20}$ cm$^2$)} & \colhead{}}

\startdata
Be & \phn73--111 & 10.5\phn & 1 \\
B & 115--188 & \phn1.7\phn & 2 \\
R1 & 110--284 & \phn1.4\phn & 3  \\
R2 & 140--284 & \phn0.86 &  3 \\
R12 & 120--284 & \phn1.0\phn &  3 \\
\enddata

\tablerefs{(1) \citealt{juda91}; (2) \citealt{mccammon83}; (3)
  \citealt{snowden97}} 
\tablenotetext{a}{10\% of peak response. Plots of the band response
  are given in Figure \ref{fig-response}.}
\tablenotetext{b}{Cross section to radiation from $\sim$$10^6$\,K
  thermal plasma.}

\end{deluxetable}
 \newpage
\begin{deluxetable}{lcccc}
\tablecolumns{5}
\tablewidth{0pt}
\tablecaption{Model Temperatures ($\log T$) for Measured Band Ratios\label{tbl-spectra}}
\tablehead{\colhead{Abundances\tablenotemark{a}} & \colhead{R1/R2} & \colhead{B/R12} &
 \colhead{Be/R12} & \colhead{Mean\tablenotemark{b}}  }

\startdata
Solar & $6.07 \pm 0.05$ & $5.78 \pm 0.08$ & $6.19 \pm 0.02$ & $6.15 \pm 0.21$ \\
Light Depletion & $5.95 \pm 0.03$ & $5.75 \pm 0.05$ & $5.87 \pm 0.02$ & $5.88 \pm 0.10$ \\
Heavy Depletion & $5.90 \pm 0.04$ & $5.74 \pm 0.04$ & $5.84 \pm 0.01$ & $5.84 \pm 0.08$ \\
\enddata

\tablecomments{Temperatures and uncertainties derived from overlap of
  \citeauthor{raymondsmith} model with measured local band ratios;
  Figure \ref{fig-bandvst}.}
\tablenotetext{a}{Abundances from \citealt{savagesembach}; see text}
\tablenotetext{b}{The mean represents the weighted average of the
  preceeding three columns while the uncertainty is the standard
  deviation of the same.}

\end{deluxetable}
 \newpage

\end{document}